\documentclass[twocolumn,showpacs,preprintnumbers,amsmath,amssymb]{revtex4}
\usepackage{graphicx}

\begin{document}
\draft
\title{Observation of dynamic localization in periodically-curved waveguide arrays} \normalsize
\author{S. Longhi, M. Marangoni, M. Lobino, R. Ramponi, and P. Laporta}
\address{Dipartimento di Fisica and Istituto di Fotonica e Nanotecnologie del CNR,
Politecnico di Milano, Piazza L. da Vinci 32, I-20133 Milano,
Italy}
\author{E. Cianci and V. Foglietti}
\address {Istituto di Fotonica e Nanotecnologie del CNR, sezione di Roma, Via Cineto Romano 42, 00156 Roma, Italy}

%\date{.}

%
\bigskip
\begin{abstract}
\noindent We report on a direct experimental observation of
dynamic localization (DL) of light in sinusoidally-curved
Lithium-Niobate waveguide arrays which provides the optical analog
of DL for electrons in periodic potentials subjected to ac
electric fields as originally proposed by Dunlap and Kenkre [D.H.
Dunlap and V.M. Kenkre, Phys. Rev. B {\bf 34}, 3625 (1986)]. The
theoretical condition for DL in a sinusoidal field is
experimentally demonstrated.

\end{abstract}

% 42.82.Et Waveguides, couplers, and arrays
% 73.23.Ad Ballistic transport
% 42.25.Bs Wave propagation, transmission and absorption

\maketitle

\newpage
The quantum motion of an electron in a periodic potential
subjected to an external field has provided since a long time a
paradigmatic model to study fascinating and rather universal
coherent dynamical phenomena. These include the long-predicted
Bloch oscillations (BO) for dc fields \cite{Bloch}, i.e. an
oscillatory motion of the wave packet related to the existence of
a Wannier-Stark ladder energy spectrum, and the more
recently-predicted dynamic localization (DL) for ac fields
\cite{Dunlap86}, in which a localized particle periodically
returns to its initial state following the periodic change of the
field. In recent years, BO have been experimentally observed in a
wide variety of systems  including semiconductor superlattices
\cite{Waschke93}, atoms in accelerated optical lattices
\cite{BEC}, and optical waveguide arrays with a transverse
refractive index gradient
\cite{Pertsch99,Morandotti99,Christodoulides03}. DL is a
phenomenon similar to BO which occurs when the electron is
subjected to an ac field. The condition for DL, as originally
predicted by Dunlap and Kenkre \cite{Dunlap86} in the
nearest-neighbor tight-binding (NNTB) approximation and for a
sinusoidal driving field $E(t)=F \sin (\omega t)$, is that
$J_0(\Gamma)=0$, where $\Gamma=eaF/ \hbar \omega$ and $a$ is the
lattice period. DL has been shown to be related to the collapse of
the quasienergy minibands \cite{Holthaus92}, and the general
conditions for DL beyond the NNTB approximation and for
generalized ac fields have been identified \cite{Dignam02}; DL
under the action of both ac and dc fields has been also studied
\cite{Zhao95}, and the influence of excitonic and many-body
effects on DL in semiconductor superlattices has been considered
(see, e.g. \cite{Zhang03,Madureira04}). Despite the large amount
of theoretical work on DL, experimental evidences of DL are very
few \cite{Keay95,Madison98} and much less persuasive than those
reported for BO. In semiconductor superlattices, the occurrence of
a variety of detrimental effects \cite{Madureira04} makes it hard
an unambiguous demonstration of DL. The observation of dynamical
Bloch band suppression for cold sodium atoms in a
frequency-modulated standing light wave has been related to DL
\cite{Madison98}, however a direct experimental evidence of DL in
real space is still lacking. Recently, it has been suggested
\cite{Lenz03,Longhi05} that optical waveguide arrays with a
periodically-bent axis may provide an ideal laboratory system for
an experimental realization of DL in optics, the local curvature
of the waveguide providing the optical equivalent of an applied
electric field \cite{Lenz99,Longhi05}. Since optical DL
corresponds to a periodic self-imaging of light \cite{Longhi05},
it bears a close connection with self-collimation and diffraction
cancellation effects that have been already demonstrated in
photonic crystals (PC) and waveguide arrays
\cite{Eisenberg00,Pertsch02,selfcollimation}. In these previous
studies, diffractionless propagation is achieved owing to the
vanishing of diffraction at the inflection points of the
isofrequency PC band surfaces \cite{Pertsch02,selfcollimation} or
by alternating the sign of diffraction \cite{Eisenberg00} using
zig-zag waveguides (diffraction management).\\
In this Letter we provide the first direct experimental
observation of DL of light in sinusoidally-curved waveguide arrays
which exactly mimics the original Dunlap and Kenkre (DK) model
\cite{Dunlap86}. The quantum-optical equivalence between light
propagation in sinusoidally-curved waveguide arrays and the motion
in a crystal of an electron subjected to an ac field has been
formally stated in Ref.\cite{Longhi05}. The effective
two-dimensional (2D) wave equation describing beam propagation in
a periodically-curved array reads \cite{Longhi05}
\begin{equation}
i \lambdabar \frac{\partial \psi}{\partial z} =
-\frac{\lambdabar^2}{2n_s} \frac{\partial^2 \psi}{\partial x^2} +
V(x-x_0(z)) \psi.
\end{equation}
where: $z$ is the paraxial propagation distance, $\lambdabar
\equiv \lambda / (2 \pi) =1/k$, $V(x) \simeq n_s-n(x)$, $n(x)$ is
the effective refractive index profile of the array with period
$a$ [$n(x+a)=n(x)$], $n_s$ is the substrate refractive index, and
$x_0(z)$ describes the periodic bending profile of the waveguide
with period $\Lambda \gg \lambda$. By means of a
Kramers-Henneberger transformation $x'=x-x_0(z)$, $z'=z$, and $
\phi(x',z')=\psi(x',z') \exp \left[-i (n_s/ \lambdabar)
\dot{x}_{0}(z') x'
   -i (n_s/ 2 \lambdabar) \int_{0}^{z'} d \xi \; \dot{x }_{0}^2(\xi)
 \right]$ (where the dot indicates the derivative with respect to
 $z'$), Eq.(1) is transformed into the Schr\"{o}dinger
 equation for a particle of mass $m=n_s$ and charge $q$ in a periodic potential
 $V(x')$ under the action of an ac field
 $\mathcal{E}(z')$:
\begin{equation}
i \lambdabar \frac{\partial \phi}{\partial z'} =
-\frac{\lambdabar^2}{2n_s} \frac{\partial^2 \phi}{\partial x'^2} +
V(x') \phi-q\mathcal{E}(z')x' \phi,
 \end{equation}
where $\lambdabar$ plays the role of the Planck constant and the
ac force is related to the axis bending by the equation
$q\mathcal{E}(z')=-n_s \ddot{x}_{0}(z')$. Note that, in the
optical analogy, the temporal variable of the quantum problem is
mapped into the spatial propagation coordinate $z'$, so that DL is
simply observed as a return of the initial light intensity
distribution during propagation. For single-mode waveguides, in
the NNTB approximation and assuming that the lowest Bloch band of
the array is excited, from Eq.(2) the following coupled-mode
equations can be derived
\begin{equation}
i {\dot c} _n= -\Delta (c_{n+1}+c_{n-1})-\frac{q n a}{\lambdabar}
\mathcal{E}(z')c_n,
\end{equation}
for the amplitudes $c_n$ of the field in the individual
waveguides, where $\Delta>0$ is the coupling constant. The
phenomenon of DL \cite{Dunlap86} corresponds to periodic
self-imaging at planes $z=0, \Lambda,2\Lambda ,3 \Lambda,...$. In
the NNTB approximation, the condition for DL is
$\int_{0}^{\Lambda} d \xi \; \exp[-i \gamma(\xi)]=0$, where
$\gamma(z)=(A n_s a/ \lambdabar)\dot{x}_0(z)$
\cite{Dignam02,Longhi05}. In particular, for a sinusoidally-bent
array with period $\Lambda$ and amplitude $A$, $x_0(z)=A \sin (2
\pi z/ \Lambda)$, the condition for DL is \cite{Dunlap86}
$J_0(\Gamma)=0$, with
\begin{equation}
\Gamma=\frac{4 \pi^2 n_s a A}{\Lambda \lambda}.
\end{equation}
The onset of DL condition $J_0(\Gamma)=0$ can be at best
visualized by a direct measure of the impulse response
[$c_n(0)=\delta_{n,0}$] of a curved array
\cite{Dunlap86,Grifoni98}. As for the straight array the light
spreads linearly with propagation distance $z$ as
$|c_n(z)|^2=|J_n(2 \Delta z)|^2$, yielding for the mean square
number of excited waveguides the diffusive law $\langle n^2
\rangle \equiv \sum_n n^2 |c_n|^2=2\Delta^2 z^2$, for the
sinusoidally-curved array, one has instead
\cite{Dunlap86,Grifoni98}
\begin{equation}
\langle n^2 \rangle=2 \Delta^2 \left[u^2(z)+v^2(z) \right],
\end{equation}
where we have set $u(z)= \int_{0}^z d \tau \cos(\mathcal{E}_0
\eta(\tau))$, $v(z)= \int_{0}^z d \tau \sin(\mathcal{E}_0
\eta(\tau))$, $\mathcal{E}_0=4 \pi^2 a n_s A / (\lambdabar
\Lambda^2)$, and $\eta(z)=[1-\cos(2 \pi z / \Lambda)]/[2 \pi /
\Lambda]$. In particular, for a modulation period $\Lambda$
smaller than the coupling length $1/\Delta$, one has $|c_n(z)|^2
\simeq |J_{n}(2 \Delta J_0(\Gamma) z)|^2$ and $\langle n^2 \rangle
\simeq 2\Delta^2 J_{0}^2 (\Gamma) z^2$. Therefore, when the
condition $J_0(\Gamma)=0$ is satisfied, an effective suppression
of waveguide coupling is attained, which corresponds to a coherent
destruction of tunneling \cite{Grifoni98}. For $J_0(\Gamma) \neq
0$, the light diffraction pattern is analogous to that of a
straight array with an effective coupling coefficient
$\Delta_{eff}=\Delta |J_0(\Gamma)|$. In our experiments, we could
vary the parameter $\Gamma$ [Eq.(4)] by either tuning the probing
wavelength $\lambda$ or by measuring, for a given wavelength, the
diffraction patterns of several arrays with different modulation
period $\Lambda$.
\begin{figure}
\includegraphics[scale=0.55]{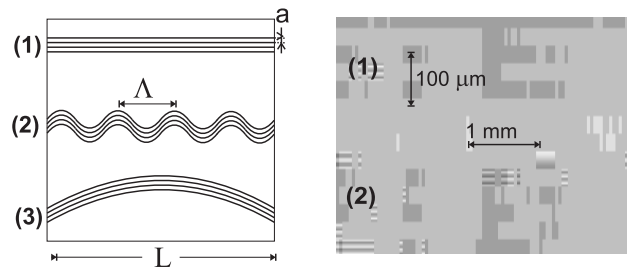} \caption{(color online)
Left: Schematic of the first manufactured sample showing the three
waveguide arrays: the straight array (1), the sinusoidally-curved
array with short period (2), and the semi-cycle sinusoidal array
(3). Right: Microscope image showing a particular of the straight
and short-period curved arrays.}
\end{figure}
\noindent The first realized sample (Fig.1) consists of a set of
one straight and two sinusoidally-curved arrays of waveguides with
a graded index profile fabricated by the annealed proton exchange
(APE) technique in $z$-cut congruent lithium niobate
\cite{Bortz91}. Channel width ($\simeq 7 \; \mu$m) and fabrication
parameters have been chosen such that the individual waveguide
turns out to be single-mode, for TM polarization, in the whole
spectral range from $\lambda=1440$ nm to $\lambda=1610$, with an
upper cut-off frequency at around $\lambda \simeq 1660$ nm. Each
array is made by a set of 80 identical and equally-spaced
$L=28$-mm-long waveguides with separation $a=14 \; \mu$m.
\begin{figure}
\includegraphics[scale=0.44]{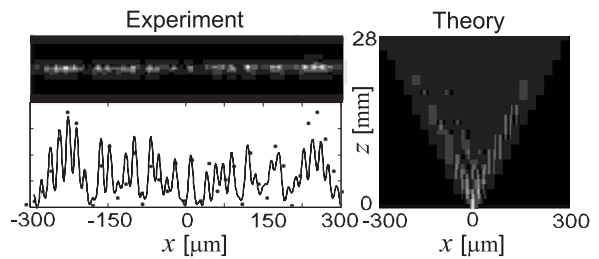} \caption{Left: 2D
output intensity pattern recorded on the IR camera for single
waveguide excitation at $\lambda=1610$ nm of the straight array,
and corresponding cross section intensity profile. The points show
the waveguide intensity distribution $|c_n|^2=|J_n(2 \Delta L)|^2$
predicted by the coupled-mode equations (3) for $\Delta=3 \; {\rm
cm}^{-1}$. Right: beam evolution (top view) predicted by numerical
simulations for the straight array.}
\end{figure}
\noindent Array (2) comprises 7 sinusoidal cycles ($\Lambda=4$ mm
and $A \simeq 13 \; \mu$m), with $\Gamma \simeq 2.405$ at $\lambda
\simeq 1610$ nm, whereas array (3) comprises only a semi-cycle
($\Lambda=56$ mm and $A \simeq 164 \; \mu$m), with $\Gamma \simeq
2.405$ at $\lambda \simeq 1440$ nm. The waveguide coupling
constant $\Delta$ has been measured as a function of the probing
wavelength $\lambda$ by the analysis of the diffraction pattern
obtained for the straight array (1) with a single waveguide
excitation at the input plane (Fig.2). Light coupling was
accomplished by focusing the circular diffraction-limited beam of
a tunable semiconductor laser (Agilent Mod. 81600B) into the input
face of one waveguide of the array to obtain a focused spot size
of radius $\simeq 3.5 \; \mu$m. The light beam distribution at the
exit of the array was imaged and recorded onto an IR Vidicon
camera.
\begin{figure}
\includegraphics[scale=0.46]{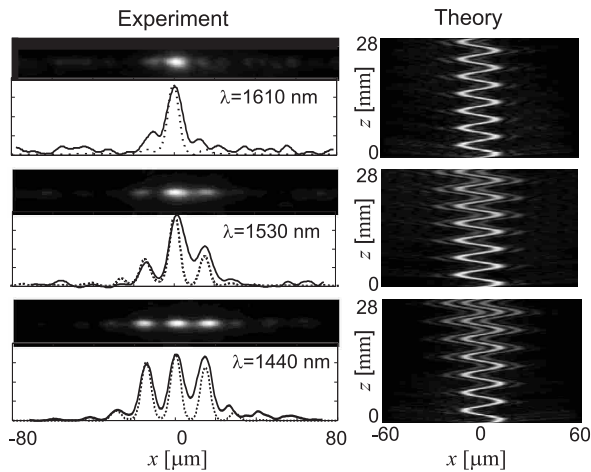} \caption{Left: 2D
output intensity patterns recorded on the IR camera for a single
waveguide excitation of the short-period curved array at different
wavelengths, and corresponding cross section intensity profiles
(solid curves). In the figures the dashed curves are the
numerically-predicted intensity profiles. Right: corresponding
beam evolution (top view) predicted by numerical simulations of
Eq.(1).}
\end{figure}
\noindent The measured coupling constant $\Delta$ increases as the
wavelength is increased and varies from $ \Delta \sim 1.75 \; {\rm
cm}^{-1}$ at $\lambda=1440$ nm to $\Delta \sim 3 \; {\rm cm}^{-1}$
at $\lambda=1610$ nm (the case shown in Fig.2). The measured
impulse response of the curved array (2) for a few values of
$\lambda$ is shown in Fig. 3, together with the beam evolution
along the arrays as predicted by a numerical analysis of Eq.(1).
\begin{figure}
\includegraphics[scale=0.46]{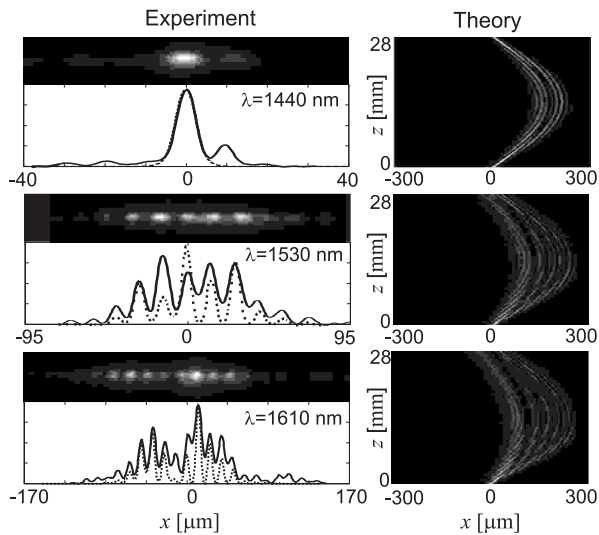} \caption{Same as Fig.3, but for
the semi-cycle curved array.}
\end{figure}
\begin{figure}
\includegraphics[scale=0.53]{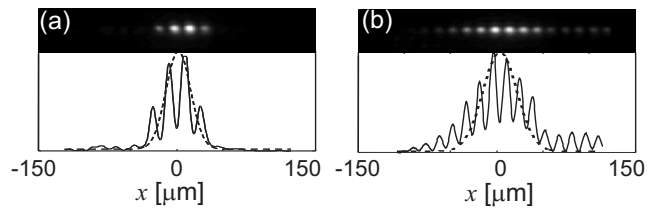} \caption{Measured 2D
output intensity patterns and corresponding cross sections for an
astigmatic Gaussian beam excitation of array (3) with (a) $w_x
\sim 24.7 \;\mu$m, and (b) $w_x \sim 37.4 \;\mu$m. The dashed
curves are the input Gaussian beam profiles in the $x$ direction.}
\end{figure}
\noindent Note that, as the wavelength $\lambda$ is tuned at the
$\Gamma=2.405$ value (top figure) a single spot is observed at the
output plane, whereas far from the DL condition  the number of
excited waveguides increases. Note however that the change of
$\Gamma$ around $\sim 2.4$ allowed by wavelength tuning is rather
modest (about $10 \%$), which explains why the number of excited
output waveguides is relatively small ($\sim 3-4$) even at
$\lambda=1440$ nm. An important property of DL with a sinusoidal
ac modulation is that wave refocusing is attained not only at
multiples of the modulation cycle as shown for array (2) in Fig.3,
but also at multiples of the semi-cycle. This is clearly
demonstrated both numerically and experimentally in Fig.4, which
shows the impulse response of array (3). We note that, in the
context of DL, the refocusing property in a semi-cycle was
theoretically investigated for an ac square-wave driving field
\cite{Lenz03,Zhu99} and related to periodic BO motion with
alternating sign for the dc field (the so-called ac BO
\cite{Lenz03}). However, the dynamics shown in Fig.4 is not
 a BO motion with a dc field, which would strictly require
circularly-curved waveguides \cite{Lenz99,Lenz03}.\\
We also checked the onset of DL by broad beam excitation
illuminating at normal incidence the input facet of the array with
an astigmatic Gaussian beam of varying width $w_x$ in the in $x$
direction. As an example, Fig. 5 shows diffraction cancellation
observed in array (3) with excitation at $\lambda=1440$ nm for two
values of $w_x$; similar results are obtained for array (2). It
should be noted that an analogous diffraction suppression was
previously observed in Ref.\cite{Eisenberg00} using zig-zag arrays
(see, in particular, Figs. 3 and 4). In that case, diffraction
cancellation arises from the alternation of positive and negative
diffraction, obtained by periodic tilting of straight waveguide
pieces. This is basically the same idea of "dispersion management"
known for temporal pulse propagation in dispersive media. This
scheme of diffraction cancellation, however, can not be regarded
to as a realization of DL, i.e. wave localization induced by a
periodic ac force, as intended within the original DK model
(sinusoidal ac field) or its generalizations \cite{Dignam02},
including a square-wave ac field \cite{Lenz03}. In fact, for
zig-zag pieces of straight waveguides the curvature $\ddot x_0$ is
zero and the $\mathcal{E}$ field vanishes in the
quantum-mechanical model [see Eq.(2)]. At the tilting planes of
waveguide axis, however, the field wave vector (momentum)
experiences an instantaneous change of its direction with respect
to waveguide axis with a corresponding change of the diffraction
sign, leading to a zero mean diffraction coefficient according to
Eq.(4) of Ref. \cite{Eisenberg00}. In the quantum mechanical model
(2), this would correspond to a quantum particle
moving in a periodic potential whose momentum is abruptly and periodically kicked.\\
To quantitatively check the diffusive law (5), we fabricated a
second sample consisting of 7 sinusoidally-curved arrays with the
same length and design parameters as those of the array (2) in
Fig.1 but with different and increasing values of $\Lambda$ [from
$2.8$ mm to $14$ mm], enabling to span a wide range of $\Gamma$
values not achievable by simple wavelength tuning. The impulse
response of the 7 arrays, excited at $\lambda=1610 \; {\rm nm}$,
was recorded, and the square mean number of excited waveguides
$\langle n^2 \rangle$ was estimated from the measured
cross-sectional profiles $|\psi(x,L)|^2$ by the relation $\langle
n^2 \rangle \sim \int dx \; (x/a)^2 |\psi(x,L)|^2$.
\begin{figure}
\includegraphics[scale=0.48]{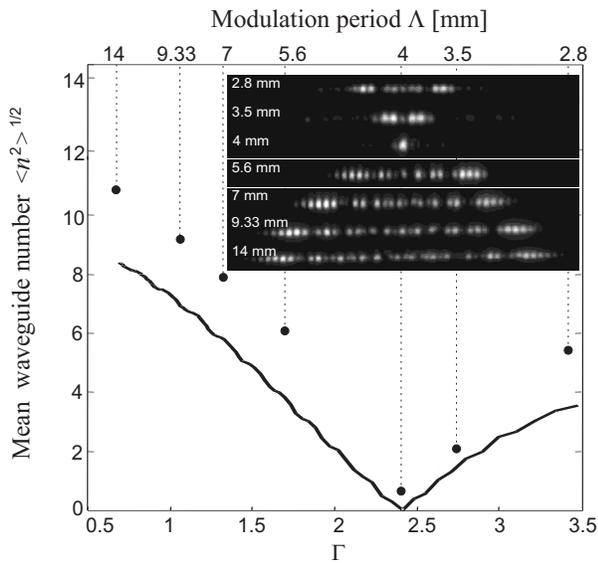} \caption{Measured
number of excited waveguides $\sqrt{\langle n^2 \rangle}$ versus
$\Gamma$ (dots) for 7 waveguide arrays with increasing values of
$\Lambda$, and corresponding theoretical behavior (solid curve).
The recorded 2D output intensity profiles are also shown.}
\end{figure}
The experimental results are summarized in Fig.6 and compared to
the theoretical prediction [Eq.(5)]. Note that the spreading of
the output patterns as $\Gamma$ departs from $2.405$ is related to
an increase of the effective coupling coefficient
$\Delta_{eff}=\Delta |J_0(\Gamma)|$, which plays an analogous role
of the mean-diffraction coefficient in the diffraction management
scheme of Ref.\cite{Eisenberg00} (compare Fig.6 with Fig.4  of
Ref.\cite{Eisenberg00}). The occurrence
of the minimum in Fig.6 is thus a clear  signature of DL.\\
In conclusion, we have experimentally observed DL of photons in
sinusoidally-curved waveguide arrays, a phenomenon which mimics
the quantum mechanical effect originally proposed by Dunlap and
Kenkre \cite{Dunlap86}. In particular, the basic condition for DL
in a sinusoidal field, $J_0(\Gamma)=0$, has been experimentally
confirmed.

\noindent This research was partially funded by MIUR (FIRB
project). The authors acknowledge A. Minotti, S. Quaresima, and M.
Scarparo for
their technical assistance.\\
\\
Author 's email address: longhi@fisi.polimi.it

%\begin{figure}[hb]
%\centerline{\scalebox{0.9}{\includegraphics{Fig3REV.EPS}}}
%\vskip3cm \caption{ S. Longhi,et al.
% Observation of ...}
%\end{figure}

\end{document}